# Plasmonic absorption characteristics based on dumbbell-shaped graphene metamaterial arrays

**Chunlian Cen[1,2], Jiajia Chen[1,2], Hang Lin[1,2], Cuiping Liang[1,2], Jing Huang[1,2], Xifang Chen[1, 2], Yongjian Tang[1,2], Zao Yi [1,2*], Xibin Xu[3*], Shuyuan Xiao[4]**

[1]*Joint Laboratory for Extreme Conditions Matter Properties, Southwest University of Science and Technology, Mianyang 621010, China*

[2] *Sichuan Civil-Military Integration Institute, Mianyang 621010, China*

[3]*Research Center of Laser Fusion, China Academy of Engineering Physics, Mianyang 621010, China*

[4] *Wuhan National Laboratory for Optoelectronics, Huazhong University of Science and Technology, Wuhan 430074, China*

∗ Correspondence should be addressed to Zao Yi, Xibin Xu

Tel: 86-0816-2480830; Fax: 86-0816-2480830

E-mail address: yizaomy@163.com; dodolong@csu.edu.cn

**Abstract:**

In this paper, we proposed a theoretical model in the far-infrared and terahertz (THz) bands, which is a dumbbell-shaped graphene metamaterial arrays with a combination of graphene nanorod and two semisphere-suspended heads. We report a detailed theoretical investigation on how to enhance localized electric field and the absorption in the dumbbell-shaped graphene metamaterial arrays. The simulation results show that by changing the geometrical parameters of the structure and the Fermi level of graphene, we can change the absorption characteristics. Furthermore, we have discovered that the resonant wavelength is insensitive to TM polarization. In addition, we also find that the double-layer graphene arrays have better absorption characteristics than single-layer graphene arrays. This work allows us to achieve tunable terahertz absorber, and may also provide potential applications in optical filter and biochemical sensing.

## 1. Introduction

Graphene, a single layer of carbon atoms in plane with a honeycomb lattice, is a kind of two-dimensional material. Because of its peculiar electrical and optical properties [1-3], there are broad application prospects in the fields of optoelectronic such as transparent electrodes [4-6], optical modulators [7-9], and photodetectors [10-13]. Moreover, graphene can support surface plasmons (SPs) in the infrared and THz window through proper doping and facilitate active plasmonic devices [14-17]. Compared with the traditional metal SPs, graphene show better performances, such as the limit of the electromagnetic energy constraints, low loss and high tunability. Graphene SPs both in theory and experiment have attracted great interest. Recently, many articles have proposed a very diverse array of graphene, such as graphene strips, disks, rings and cross structures [18-22]. Plasmonic metamaterial structures have been very attractive in recent years due to their unique active tunability [23,24].

In the far infrared and THz regimes, graphene has a strong plasmonic response. The localized surface plasmon resonances (LSPR) is excited in patterned graphene structures, which greatly enhances the plasmonic absorption characteristics [18,25-27]. In the near-infrared or visible regime, metal plasmonic structures [28-30], photonic crystals [31-33] or optical cavities [34-36] are introduced to enhance the near field resonance of the neighboring graphene, which increases the strength of the interaction between graphene and the incident electromagnetic wave, thereby improving absorption. Strong resonance caused strong absorption enhancement, but a narrow bandwidth, which is harmful in broadband applications. The design of different sizes of plasmonic elements can partially expand the bandwidth [37,38]. However, there is still limitations such as narrowband for current THz absorbers. We proposed that the dumbbell-shaped graphene metamaterial arrays can obtain the ideal optical properties through the configuration of geometrical parameters, beyond the intrinsic properties of the material.

In this research, we introduce the dumbbell-shaped graphene metamaterial arrays and study on its plasmonic absorption characteristics. Through the finite difference time domain (FDTD) method the absorption characteristics of arrays are numerically calculated. Compared with other shapes, geometric parameters of the dumbbell-shaped structure has a greater freedom, can be more flexible to adjust the absorption. Changing the geometrical parameters of the dumbbell-shape also has a great impact on the absorption characteristics. It is also encouraging that, under TM polarization, the absorption characteristics of arrays show excellent stability. The dynamic tunability of Fermi level is also studied. Through the study of the system, the mechanism of the absorption of electric field distribution and charge density distribution is revealed. The high absorption rate is due to the near-field enhancement caused by the surface plasmonic mode on the interface. The absorption for double-layer graphene arrays is also researched. The proposed absorber in plasmonic sensors, modulator and all-optical switch has potential application prospects.

## 2. **The geometric structure and numerical model**

Figure 1 shows the structure we proposed, consist of graphene nanorod and semispheres constructing a dumbbell-shaped graphene metamaterial arrays with period *a*, length *L*, width *W* and radius *R* of two semisphere-suspended heads. The graphene array is attached to the silicon substrate and dissociated by a thin $SiO_2$ filler piece with a thickness *d*. The incident Angle from the air to the plane is *θ*. The incident plane is the *x*-*z* plane.

The graphene conductivity is derived using the random-phase approximation (RPA) in the local limit, including both the intraband and interband processes [39,40]

$$\sigma_{\mathrm{g}} = \sigma_{\mathrm{intra}} + \sigma_{\mathrm{inter}} = \frac{2e^2 k_B T}{\pi\hbar^2}\frac{i}{\omega + i\tau^{-1}}\ln[2\cosh(\frac{E_F}{2k_BT})] + \frac{e^2}{4\hbar}[\frac{1}{2} + \frac{1}{\pi}\arctan(\frac{\hbar\omega - 2E_F}{2k_BT}) - \frac{i}{2\pi}\ln\frac{(\hbar\omega + 2E_F)^2}{(\hbar\omega - 2E_F)^2 + 4(k_BT)^2}] \tag{1}$$

where *e* is the charge of an electron, $k_B$ is the Boltzmann constant, *T* is the operation temperature, $\hbar$ is the reduced Planck's constant, $\mu$ is the angular frequency of the incident light, $\tau$ is the carrier relaxation time and $E_F$ is the Fermi level.

In the lower THz range, the interband contributions can be completely ignored. Because of Pauli exclusion principle, the surface conductivity can be approximated as a Drude-like model [41,42].

$$\sigma_g = \frac{e^2 E_F}{\pi\hbar^2} \frac{i}{\omega + i\tau^{-1}} \tag{2}$$

Here in the relaxation time $\tau = (\mu E_F)/(e v_F^2)$ depends on the the electron mobility μ = 10000 $cm^2$/V s, the Fermi level $E_F$ and the Fermi velocity $v_F = 10^6$ m/s.

The simulations are based on the FDTD method, which is a famous method for solving Maxwell's equations in time domain. The equation for Maxwell's source free zone are given as [43]

$$\nabla \times E = -\mu \frac{\partial H}{\partial t} \tag{3}$$

$$\nabla \times H = \varepsilon \frac{\partial E}{\partial t} \tag{4}$$

where E and H are the electric and magnetic fields, respectively. $\varepsilon$ is the permittivity and $\mu$ is the permeability of the medium. In FDTD method, Maxwell's equations are quantized based on the time and space. The method of solving Maxwell's equations is based on the material parameters and initial conditions, and calculates the electromagnetic field number at each time point. The incident electric field (E) is in the x direction and the plane of incidence is x-z. The absorption is calculated by A=1-T-R formula, where T and R represent the transmission and reflection, respectively.

**3. Results and discussions**

Under the same geometrical parameters and conditions, we studied the absorption characteristics of rectangle, nanorod, circle and dumbbell-shaped. Through FDTD simulation calculation, we can obtain the absorption maximum of the rectangle is 0.129, the absorption maximum of the nanorod is 0.133, the absorption maximum of the circle is 0.127, and the absorption maximum of the dumbbell-shaped is 0.142 as shown in Fig. 2(A). By comparing the absorption maximum of these four structures, it is obvious that the absorption characteristics of the dumbbell-shaped is better than those of the other three structures. Therefore, the following work is mainly to study the absorption characteristics of the dumbbell-shaped structure by changing some geometric parameters of the dumbbell-shaped structure and Fermi level of graphene. In Fig. 2(B), we have plotted the electric field distributions of four structures, and it can be seen that the electric field distribution is concentrated on the edges of their arms.

In the development of photonics and metamaterials, the most encouraging aspect is that it can obtain ideal optical properties through the configuration of geometric parameters, beyond the intrinsic properties of materials. For the periodic structure of micro-nano scale, the effective optical characteristics are the parameters of high dependence on space ratio, period, and radius. As is mentioned above, with the width-to-radius ratio W/R and length-to-period ratio L/a of the dumbbell-shaped graphene metamaterial arrays varies, there are always clear absorption peaks in the far infrared and THz ranges, as shown in Fig. 3. In fig. 3(A) shows the width *W* changes from 0.3 to 0.8 μm the incidence of a plane wave in $\theta = 0°$. With the width *W* increasing from 0.3 to 0.8 μm with an interval of 0.1 μm , and the corresponding the resonant peak undergo a blue shift. The absorption maximum increased from 0.123 to 0.176. As the width *W* gradually approaches the radius *R*, the dumbbell-shaped gradually changes to the nanorod. Under the circumstances, the absorption maximum of the nanorod graphene arrays may be greater than that of the dumbbell-shaped graphene metamaterial arrays. However, the dumbbell-shaped graphene metamaterial arrays have additional degrees of freedom in geometry. By changing geometric parameters, more free adjustable absorption characteristics can be obtained. Different width of the absorption peak of the electric field distribution (|E|), as shown in figure 3(B). The research shows that the strong position of electromagnetic field can be realized, while the electric field is mainly concentrated on the edge of the two semisphere-suspended heads. This phenomenon is caused by a strong electric dipole resonance, which is caused by the accumulated charge on the edge of the two semisphere-suspended heads. This strong resonance has effectively captured the light energy, and has provided enough time to eliminate it through the Ohmic losses in graphene.

In Fig. 3(C), we studied the length-to-period ratio L/a, when the period was fixed to $a$ = 2.5 μm. By equations $\lambda_m = 2\pi n_{sp} L/(\pi - \delta)$ , where $n_{sp}$ is the effective index of SPPs and $\lambda_m$ the resonance wavelength. The $\lambda_m$ increases with the increase of $L$. Therefore, as the length $L$ increase, the peak experiences a red shift, and the absorption maximum gradually increases. The resonant wavelength shifts from 66 μm to 93 μm and absorption maximum increases from 0.115 to 0.165. With the increase of $L$, the electric field is distribution concentrated on the edge of the two semisphere-suspended heads, as shown in Fig. 3(D).

The effect of $R$ on the absorption spectra of the proposed structure was also investigated. The absorption spectra of the dumbbell-shaped graphene metamaterials arrays by changing simultaneously the two semisphere-suspended heads radius from 0.2 μm to 0.6 μm are presented in Fig. 4(A). The resonant wavelength first experiences a red shift and then undergoes a blue shift. In Fig 4(B) the electric field is distribution concentrated on the edge of the two semisphere-suspended heads. This indicates that the resonator of the dumbbell-shaped has a strong coupling to the electric field. The effective index of SPPs $n_{sp}$ depends largely on the radius $R$ of semisphere-suspended head. As R<<L, the more narrow semisphere-suspended head suggests larger $n_{sp}$. In view of x-polarization, the semisphere-suspended head radius $R$ is much smaller than the length $L$ of the graphene nanorod, so the $n_{sp}$ of the graphene nanorod is much smaller than the semisphere-suspended head. The light tends to focus on the larger $n_{sp}$, therefore, the electric field distribution is concentrated on the edge of the semisphere-suspended head.

Figure 5(A) shows the absorption spectra at different periods. It is obvious that the period has a great influence on the absorption maximum, but the effect on the resonant wavelength is small. The resonant wavelength and absorption maximum of different periods are shown in Fig. 5(B). This indicates that the resonant wavelength is first reduced as the period increases and then hovering around 83 μm. Meanwhile, the absorption maximum reduces. The physical mechanism is that with the period increases, the resonant wavelength is nearly constant, because the resonance conditions don't change much. The filling coefficient of graphene decreases, resulting in less absorption. Because of the coupling between the adjacent nanorod and the two semispheres, the resonant wavelength has a slight change in the small period. With the period changes, the electric field is always concentrated on the edge of the two semisphere-suspended heads, as shown in fig. 5(A) is inserted.

Figure 6(A) reveals the relationship between absorption spectra and polarization under different incident angles. The resonant wavelength and absorption maximum of different $\theta$ are shown in Fig. 6(B). For TM polarization, it indicates that as $\theta$ increases, the resonant wavelength remains at 83 μm, while the absorption maximum increases significantly. The physical origin of this phenomenon is that the electric dipole resonances between the adjacent unit cells in x directions is excited by the x-polarized TM incident THz waves. The insert in Fig. 6(A) shows the electric field distributions ($|\mathbf{E}|$) at the absorption peak for TM polarization. It turns out that the electric field is distributed along the edge of the two semisphere-suspended heads and along the direction of the incident field.

The optical properties of graphene are highly dependent on its Fermi level, which is the source of the dynamic tunability of SPR based on graphene metamaterials. In order to show that the spectral tunability of the dumbbell-shaped graphene metamaterial arrays we proposed, figure 7 illustrates the absorption characteristics of graphene light-absorbing material when four structures are with different Fermi levels. As shown in Fig. 7(A), the resonant wavelength shifts from 48 μm to 39 μm and absorption maximum increases from 0.100 to 0.165 for the rectangle. In Fig. 7(B), the resonant wavelength shifts from 56 μm to 46 μm and absorption maximum increases from 0.107 to 0.180 for the nanorod. The resonant wavelength shifts from 30 μm to 24 μm and absorption maximum increases from 0.098 to 0.156 for the circle, as shown in Fig. 7(C). In Fig. 7(D), the resonant wavelength shifts from 83 μm to 68 μm and absorption maximum increases from 0.120 to 0.211 for the dumbbell-shaped. By comparing the absorption values of these four structures, the absorption properties of the dumbbell-shaped are the best. The common phenomenon of four different structures is that with the increase of Fermi level, the resonant wavelength has experienced a significant blue shift, while the absorption maximum increases at the same time. The reason is that because of the increase in Fermi level of graphene, the wavelength of SPR (fixed frequency or vacuum wavelength), rectangular, nanorod, circle and dumbbell-shape resonators are relatively small compared to the incident light, so the resonance experience is obviously blue shift. At the same time, the increase in Fermi level also results in less loss of surface conductivity of the graphene and SPR, thus causing an increase in the number of charge carriers that resonate. As a result, the field enhancement with the higher $E_F$ is stronger than with the lower $E_F$, which leads to a higher absorption strength in the light-absorbing material. Therefore, by controlling the Fermi level of the graphene, the plasmonic metamaterial structures to achieve the tunable light trapping and absorption enhancement. We would like to express here that the dumbbell-shaped graphene metamaterial arrays resonator is better at different Fermi levels, it has better absorption characteristics, which is easier to manufacture and operate than previous studies.

The application of graphene arrays with double-layer structure has better absorption than single-layer graphene arrays [27,44]. In this paper, we proposed a double-layer dumbbell-shaped graphene arrays. Fig. 8(A) describes the structure of double-layer graphene arrays separated by $SiO_2$ layer with thickness *h*. Other parameters are unchanged (*L* = 1.2 μm, W = 0.4 μm, *R* = 0.4 μm, *a* = 2.5 μm, *d* = 0.3 μm). In Fig. 8(B), the absorption peaks of monolayer and double layers are 83.1 μm and 101.9 μm, respectively. We can clearly see that the absorption of the double layers is greater than that of the single layer. The increase of the number of layers can enhance the coupling between light and graphene and increase the absorption of graphene. The absorption spectra of the rectangle for different thickness *h* double layers graphene in Fig.8(C). With the increase of *h*, the resonant wavelength exhibits a red shift and the maximum absorption is almost unchanged. As the distance between the two graphene arrays decreases, the coupling between them becomes stronger.

**4. Conclusions**

In conclusions, a plasmonic absorption characteristics based on dumbbell-shaped graphene metamaterial arrays have been designed and theoretically demonstrated. Through the simulation calculation, we found that the dumbbell-shaped graphene metamaterial arrays have better absorption properties than the rectangles, the nanorods and the circles. In our main research dumbbell-shaped graphene metamaterial arrays, the absorption properties can be regulated by changing the geometric parameters of the structure. Such as, when the *W* increases, the absorption peaks appear blue shift. However, when the *L* increases, the absorption peaks appear red shift. When the *R* is increased from 0.2 μm to 0.6 μm, the resonant wavelength first experiences a red shift and then undergoes a blue shift. When the period *a* is decreased from 4.0 μm to 2.3 μm, the maximum absorption is increased. For TM polarization, the incident angle $\theta$ do not affect the resonant wavelength, while have a great impact on the absorption peak. We also studied the absorption characteristics of these four structures (dumbbell-shaped, rectangle, nanorod and circle) by changing Fermi level of graphene. In particular, through simple operation of graphene Fermi level, at a particular wavelength can be effectively absorbed, this in the mid-infrared and detect terahertz spectrum and polarization selectivity has great potential application value. In addition, we have studied the double-layer graphene arrays, and we have found that the double-layer absorption peak red shift is better than the monolayer. With dumbbell-shaped graphene metamaterial arrays has a good tunability, we believe that this kind of metamaterial has great application prospect in future practical applications.

**Acknowledgements**

The work is supported by the National Natural Science Foundation of China (No. 51606158; 11604311), the Funded by Longshan academic talent research supporting program of SWUST (No. 17LZX452).

**Additional information**

Competing financial interests: The authors declare no competing financial interests.

**Figure captions:**

**Figure 1.** The schematic diagram of geometric structure is designed as follows: the dumbbell-shaped graphene metamaterial arrays with period *a*, length *L* and a width *W* and radius *R* of two semisphere-suspended heads. The arrays are supported on a Si substrate coated by a thin $SiO_2$ layer with thickness *d*. The incident angle is *θ*.

**Figure 2.** (A) The absorption spectra of the four structures of rectangle, nanorod, circle and dumbbell-shaped in the same conditions. (B) The electric field distribution of four different structures at absorption peak.

**Figure 3.** (A) The absorption spectra of graphene with different widths (*W*). Other parameters are unchanged (*L* ) = 1.2 μm, *R* = 0.4 μm, *a* = 2.5 μm, *d* = 0.3 μm). The *W* increases from 0.3 μm to 0.8 μm. (B) The electric field distributions for the structure of *L* = 0.3, 0.4, 0.5, 0.6 μm. (C) The absorption spectra of graphene with different lengths (*L*). Other parameters are unchanged (*W* = 0.4 μm, *R* = 0.4 μm, *a* = 2.5 μm, *d* = 0.3 μm). The *L* increases from 0.8 μm to 1.6 μm. (D) The electric field distributions for the structure of *L* = 1.0, 1.2, 1.4, 1.6 μm.

**Figure 4.** (A) The absorption spectra of graphene with different radii (*R*). Other parameters are unchanged (*L* = 1.2 μm, *W* = 0.4 μm, *a* = 2.5 μm, *d* = 0.3 μm). The *L* increases from 0.2 μm to 0.6 μm. (B) The electric field distributions for the structure of *W* = 0.2, 0.3, 0.4, 0.5 μm.

**Figure 5.** (A) The absorption spectra of graphene with different periods (*a*). (B) Dependence of the resonant wavelength (red curve) and absorption maximum (blue curve) on the period. The insert shows distributions of electric field (|E|) for (A). In (A) and (B) other parameters are unchanged (*L* = 1.2 μm, *W* = 0.4 μm, *R* = 0.4 μm, *d* = 0.3 μm). The *a* increases from 2.3 μm to 4.0 μm.

**Figure 6.** (A) The Absorption spectra of TM polarization under different incident angles (*θ*). The insert shows the electric field distributions (|**E**|) at $\theta = 0°$. (B) The resonant wavelength (red curve) and absorption maximum (blue curve) at various incident angles for TM polarization. Other parameters are unchanged ($L$ = 1.2 μm, $W$ = 0.4 μm, $R$ = 0.4 μm, $a$ = 2.5 μm,$d$ = 0.3 μm). The $\theta$ increases from 0° to 60°.

**Figure 7.** (A) The absorption spectra of the rectangle for different Fermi energy levels. (B) The absorption spectra of the nanorod for different Fermi energy levels. (C) The absorption spectra of the circle for different Fermi energy levels. (D) The absorption spectra of the dumbbell-shaped for different Fermi energy levels. Other parameters are unchanged ($L$ = 1.2 μm, $W$ = 0.4 μm, $R$ = 0.4 μm, $a$ = 2.5 μm, $d$ = 0.3 μm). $E_F$ increases from 0.6 eV to 0.9 eV.

**Figure 8.** (A) The structure consisting of double layers of dumbbell-shaped graphene metamaterial arrays. (B) Dependence of the resonant wavelength and absorption maximum on the period. (C) The absorption spectra of the rectangle for different thicknesses ($h$). Other parameters are unchanged ($L$ = 1.2 μm, $W$ = 0.4 μm, $R$ = 0.4 μm, a = 2.5 μm, $d$ = 0.3 μm). The $h$ increases from 0.1 μm to 0.5 μm.

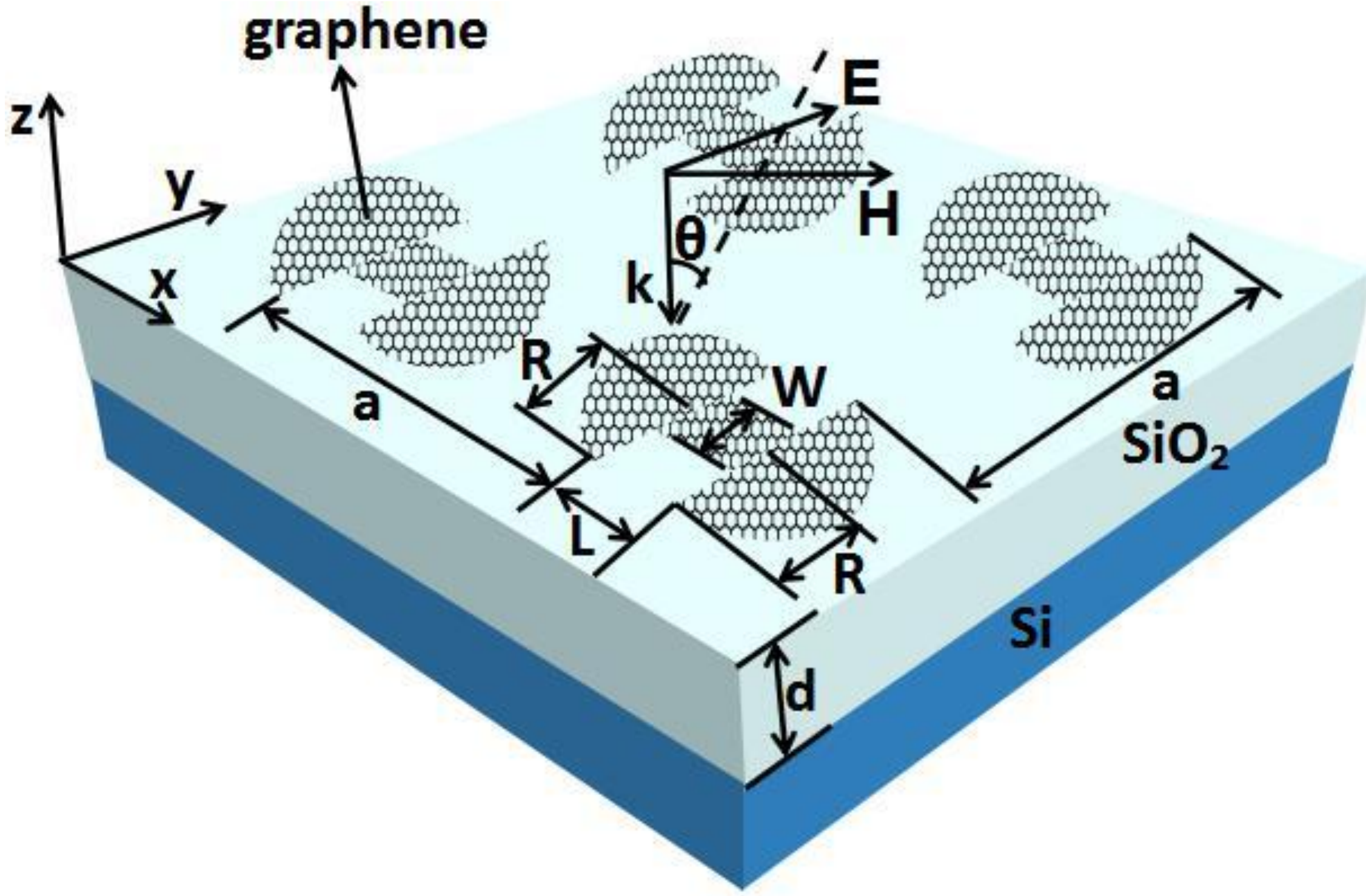

Figure 1

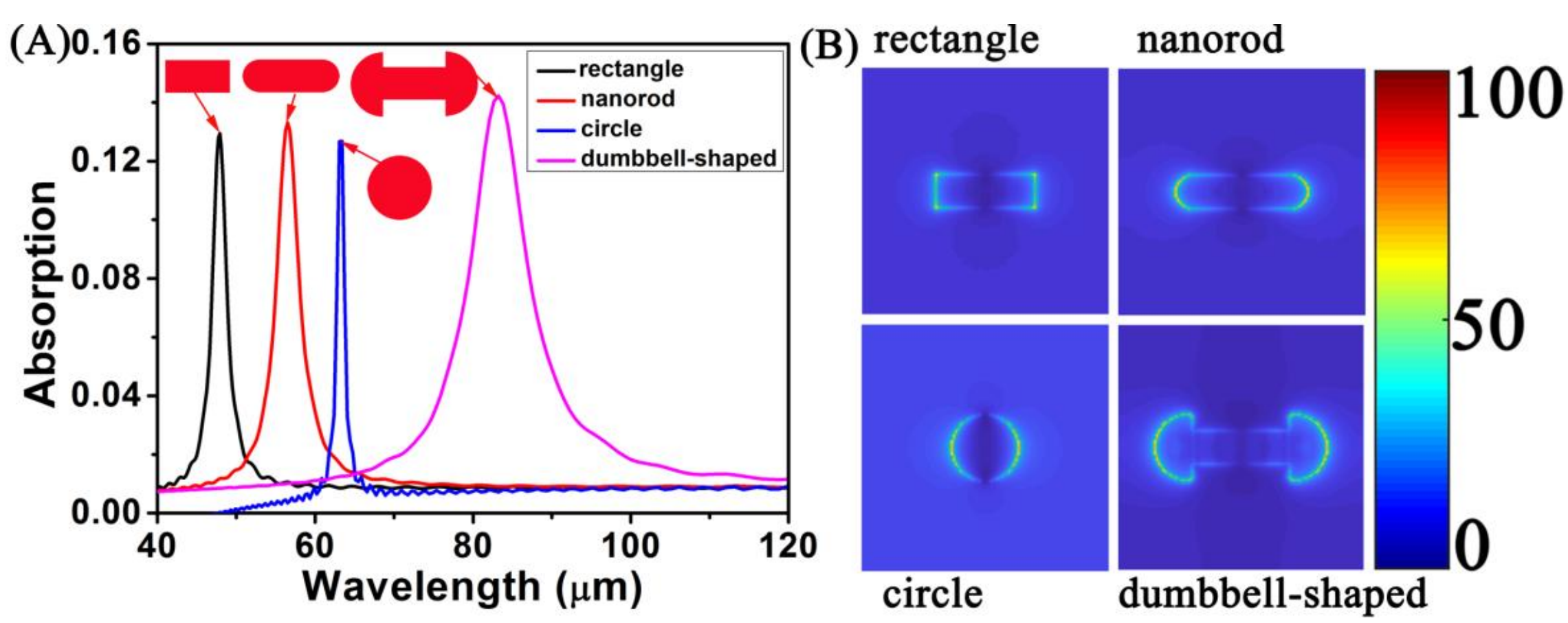

Figure 2

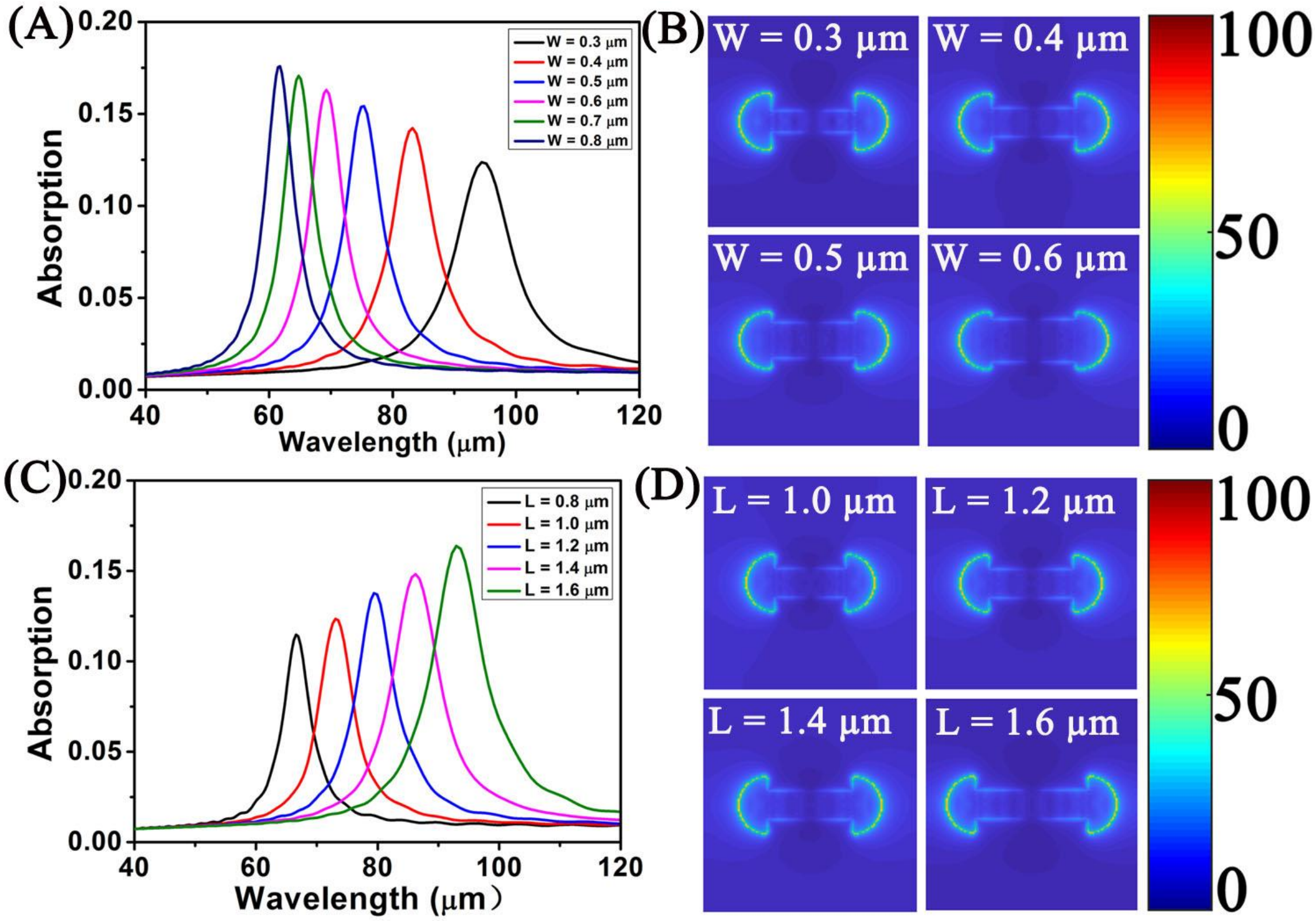

Figure 3

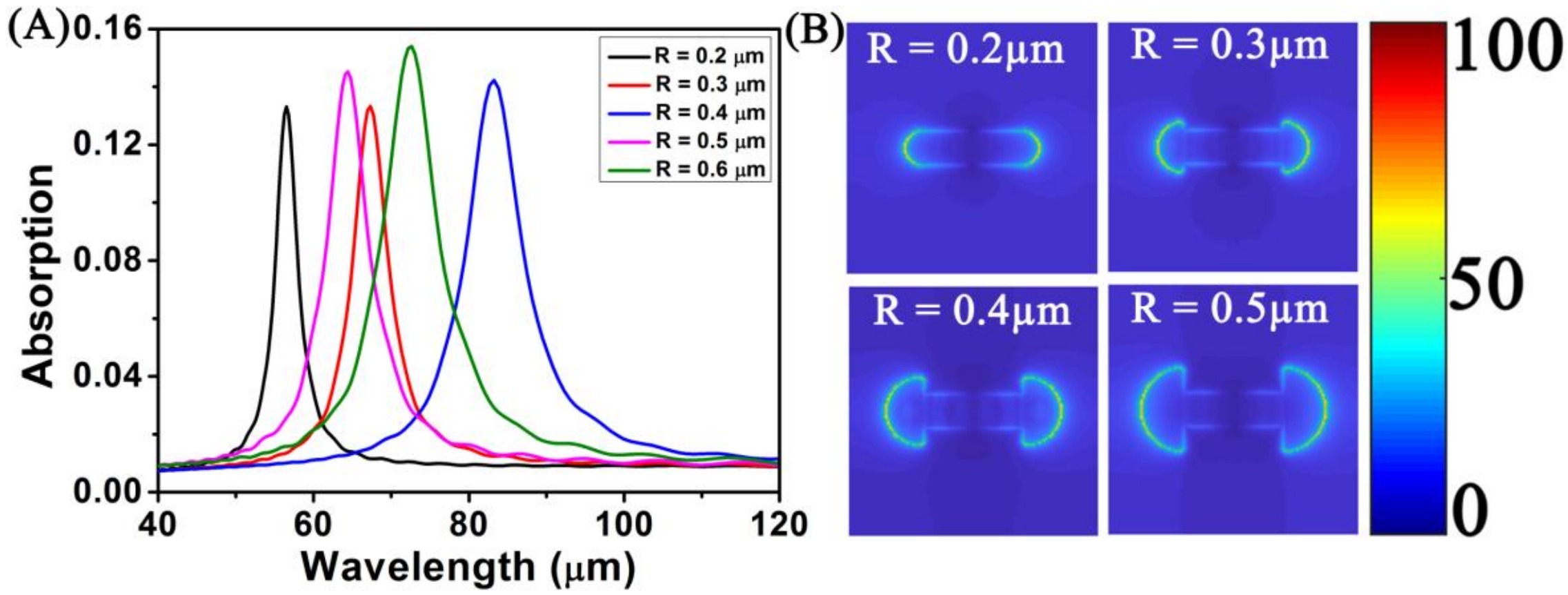

Figure 4

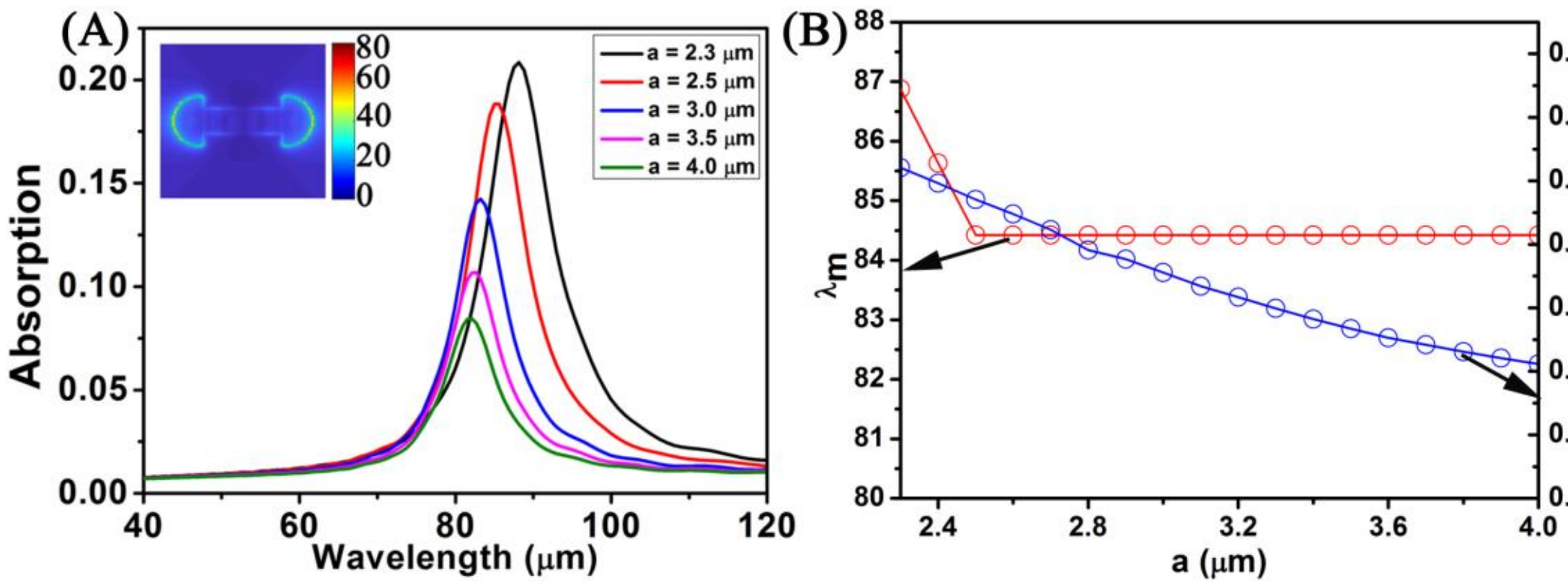

Figure 5

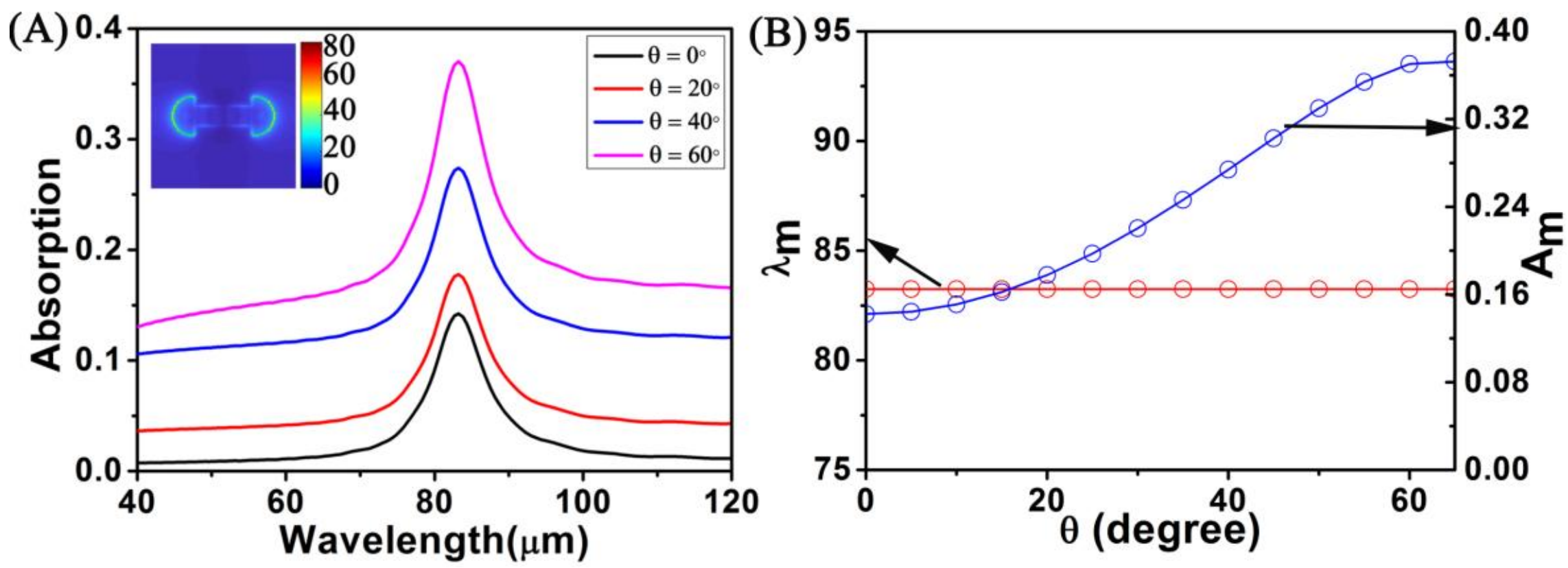

Figure 6

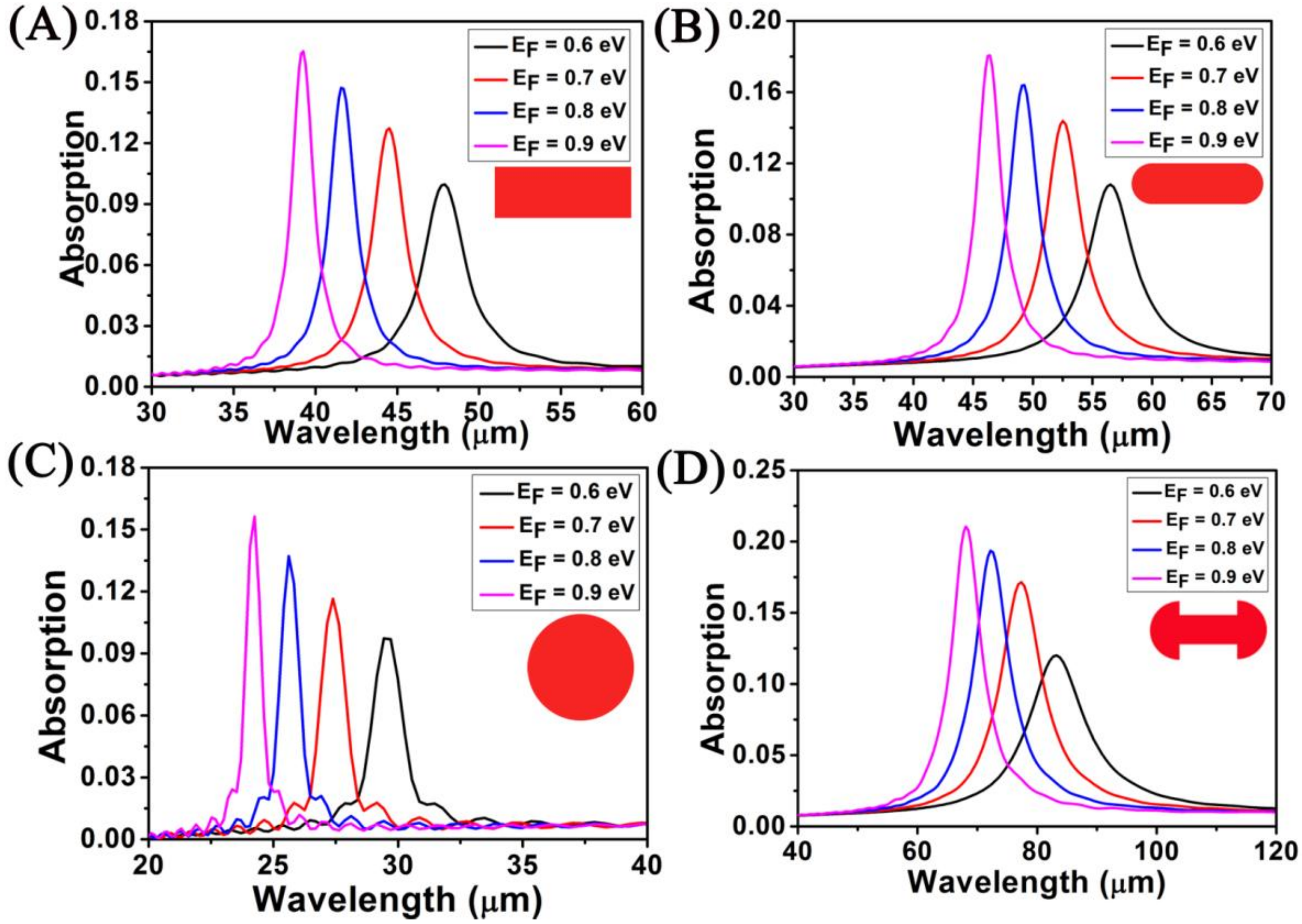

Figure 7

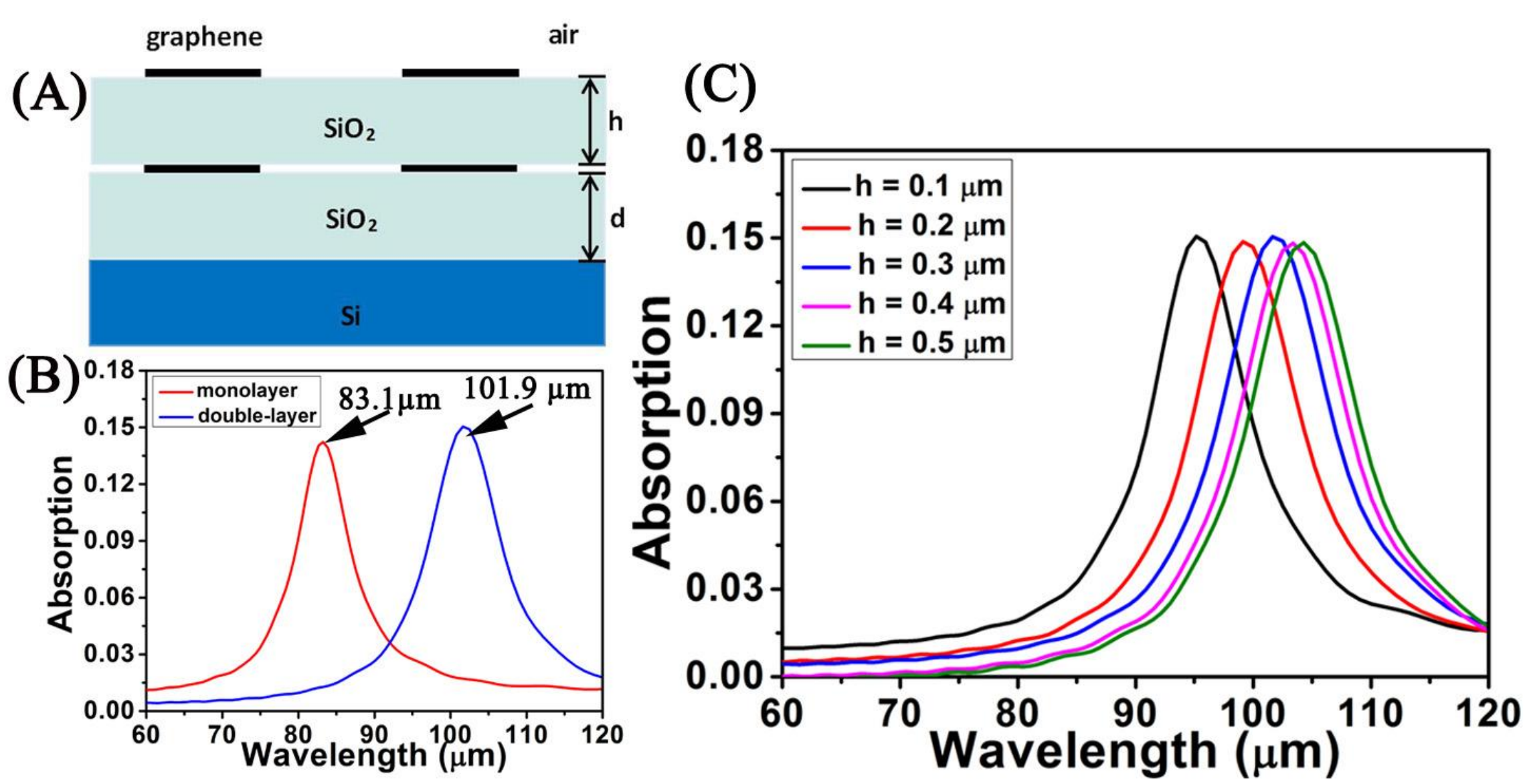

Figure 8